\documentclass[a4paper,12pt]{article}
\usepackage{amsthm}
\usepackage{amssymb}
\usepackage{amsmath}
\usepackage{amsfonts}
\usepackage{color}
\usepackage{graphicx}
\usepackage[square, comma, sort&compress, numbers]{natbib}

\newtheorem{remark}{Remark}[section]

\newtheorem{lemma}{Lemma}[section]
\newtheorem{theorem}{Theorem}[section]

\newtheorem{definition}{Definition}[section]

\def\b1{\mbox{\boldmath $1$}}

\newenvironment{demo*}{\vspace{3mm}\noindent{\bf Proof.}}{\hfill $\Box$ \vspace{3mm}}

\begin{document}
\title{\bf \Large {New class  of distortion risk measures and their tail asymptotics with emphasis on VaR}}
{\color{red}{\author{
\normalsize{Chuancun Yin\;\; Dan Zhu}\\
{\normalsize\it  School of Statistics,  Qufu Normal University}\\
\noindent{\normalsize\it Shandong 273165, China}\\
e-mail:  ccyin@mail.qfnu.edu.cn}}}
\maketitle
\vskip0.01cm
\noindent{\large {\bf Abstract}}  { {Distortion risk measures are  extensively  used   in finance and insurance applications because of their appealing properties.  We present three methods to construct
new class of  distortion functions and measures. The approach involves the  composting methods, the  mixing methods and the  approach that
based on the theory of copula.  Subadditivity is an important property when aggregating risks in order to preserve the benefits of diversification. However, Value at risk (VaR), as the  most well-known example  of  distortion risk measure   is not always globally subadditive,   except  of elliptically distributed risks. In this paper, instead of study subadditivity we investigate the tail subadditivity    for VaR and other  distortion risk measures.
In particular, we demonstrate that  VaR is tail subadditive  for the case where   the support of risk is bounded.
Various examples   are also presented to illustrate the results. }}

\medskip


\noindent{{\bf Mathematics Subjection Classification}   62P05  $\cdot$ 91B30}



\numberwithin{equation}{section}
\section{Introduction}\label{intro}
A risk measure $\rho$ is a mapping from the set of random variables
$\mathcal{X}$, standing for risky portfolios of assets and/or liabilities,  to the real line $\Bbb{R}$.
 In the subsequent discussion, positive values
of elements of $\mathcal{X}$ will be considered to represent losses, while negative values will represent gains.
 Distortion risk measures are    a particular and most important family of risk
measures that have been extensively used in finance and insurance as capital requirement and principles of
premium calculation for the regulator and supervisor. Several popular risk
measures belong to the family of distortion risk measures. For example, the value-at-risk (VaR), the tail value-at-risk (TVaR) and the Wang distortion measure.
Distortion risk measures satisfy a set of properties including
positive homogeneity, translation invariance and monotonicity.
When the associated distortion function is concave, the distortion
risk measure is also subadditive (Denneberg, 1994; Wang and Dhaene, 1998; Wirch and Hardy, 2001). VaR is one of the most popular risk measures  used in  risk management and banking supervision due to its computational simplicity and for some regularity reasons, despite has some shortcomings as a risk measure. For example, VaR is not a subadditive risk measure (see, for instance, Artzner et al. (1999), Denuit et al., (2006)),
it only concerns about the frequency of risk, but not the size of risk.
TVaR, although being coherent,  concerns only losses exceeding the VaR  and ignores useful information of the loss distribution below VaR.
Clearly, it is difficult to believe that a unique risk measure
could capture all  characteristics of risk, so that an ideal measure does not exist. Moreover,
 since risk measures associate a single number to a risk, as a matter of fact, they
cannot  exhaustively all the    information of  a risk.  However, it is reasonable to search for risk measures which are ideal for
the particular problem under investigation. As all the proposed
risk measures have drawbacks and limited applications, the selection of the appropriate  risk measures continues to be a hot topic in risk management.

Zhu and Li (2012) introduced and studied the tail distortion risk measure  which was reformulated by Yang (2012) as
follows. For a distortion function $g$, the tail distortion risk measure at level $p$ of a loss variable $X$ is defined as
the distortion risk measure with distortion function
\begin{eqnarray*}
  g_p(x)=\left\{  \begin{array}{ll} g\left(\frac{x}{1-p}\right),  \ &{\rm if}\ 0\le x\le 1-p,\\
1, \ &{\rm if}\ 1-p< x\le 1. \end{array}
  \right.
\end{eqnarray*}
Some properties and applications can be found in  Mao, Lv and Hu (2012), Mao and Hu (2012) and Lv, Pan and  Hu (2013).

As an extension of VaR and TVaR, Belles-Sampera et al. (2014a) proposed a class of new distortion risk measures called
GlueVaR risk measures, which   can be expressed as a combination of
VaR and TVaR measures at different probability
levels. They obtain the analytical closed-form expressions for the most frequently used distribution functions in
financial and insurance applications, while a subfamily of these risk measures has been shown to satisfy the
tail-subadditivity property which means that the benefits of diversification can be preserved, at least they hold in extreme cases.
 The applications of GlueVaR risk measures in capital allocation  can be found in the recent paper Belles-Sampera et al. (2014b).

Cherubini and Mulinacci (2014) propose a class of distortion measures based on contagion from an external ``scenario" variable. The dependence between the scenario and the variable whose risk is measured is modeled with a copula function with horizontal concave sections, they give conditions to ensure that coherence requirements be met, and propose examples of measures in this class based on copula functions.

The first purpose of this paper is to construct new risk measures  following  Zhu and Li (2012), Belles-Sampera et al. (2014a) and Cherubini and Mulinacci (2014).  The newly introduced risk measures  are included the tail distortion risk measure   and the GlueVaR  as specials.
The second goal of the paper is to  investigate the tail asymptotics of distortion risk measures  for the sum of possibly
dependent risks with emphasis on VaR.
The rest of the paper is organized as follows. We review some basic definitions and notations  such as distorted functions, distorted expectations and distortion risk measures in Section 2. In Section 3  several  new  distortion functions and risk measures are introduced.
In Section 4 we investigate  the tail asymptotics as well as subadditivity/superadditivity of VaR.
Finally, in Section 5 we analyze the subadditivity  properties of a class of distortion risk measures.

 \vskip 0.2cm
 \section{Distortion risk measures}
\setcounter{equation}{0}

\subsection {Distorted  functions}

A distortion function is a non-decreasing function
$g:[0,1]\rightarrow [0,1]$   such that
$g(0)=0, g(1)=1$. Since Yaari (1987) introduced  distortion function in  dual theory of choice under
risk, many different distortions $g$ have been proposed in the literature.
Here we list some commonly used distortion functions.   A
summary of other proposed distortion functions can be found in Denuit et al. (2006).

$\bullet$ $g(x)={\bf 1}_{(x>1-p)}(x)$,
where  the notation ${\bf 1}_{A}$ to denote the indicator function, which equals 1 when $A$ holds true and 0 otherwise.


$\bullet$ $g(x)=\min\{\frac{x}{1-p},1\}$.

$\bullet$ Incomplete beta function $g(x)=\frac{1}{\beta(a,b)}\int^x_0t^{a-1}(1-t)^{b-1}dt$, where $a>0$ and $b>0$ are parameters and
 $\beta(a,b)=\int^1_0 t^{a-1}(1-t)^{b-1}dt$. Setting $b=1$ gives the power distortion $g(x)= x^{a}$. Setting $a=1$
gives the dual-power distortion $g(x)= 1-(1-x)^{b}.$

$\bullet$  The Wang distortion $g(x)=\Phi(\Phi^{-1}(x)+\Phi^{-1}(p)), 0<p<1,$
where $\Phi$ is the distribution function of the standard normal.

$\bullet$ The  lookback distortion  $g(x)=x^p(1- p \ln x),  p\in (0, 1].$

Obviously,   every concave distortion function is continuous on
the  interval $(0, 1]$ and  can have jumps in 0. In contrast, every convex distortion function is continuous on
the  interval $[0, 1)$  and  can have jumps in 1. For a distortion function  $g$, if there exists a $t_0>0$ such that $g(t_0)=0$, then
$g$ is not  concave; if there exists a $t_1<1$ such that $g(t_1)=1$, then
$g$ is not  convex. The identity function is the smallest concave distortion function and also the  largest convex distortion function;
$g_0(x):={\bf 1}_{(x>0)}$ is concave on $[0,1]$ and is  the largest distortion function.  $g^0(x):={\bf 1}_{(x=1)}$ is convex on $[0,1]$ and is  the smallest distortion function.
For $0<p<1$, we remark that
$g_1(x):=\min\{\frac{x}{1-p},1\}$ is the smallest concave distortion function such that $g_1(x)\ge {\bf 1}_{(x>1-p)}(x)$. In fact, we consider a concave distortion function $g$  such that $g(x)\ge {\bf 1}_{(x>1-p)}$,
then $g\equiv 1$ on $(1-p,1]$. As $g$ is concave, it follows that
$g(x)\ge \frac{x}{1-p}$ for $x\le 1-p$, and thus $g(x)\ge \min\{\frac{x}{1-p},1\}$ for $0<x<1$.
Any concave distortion function $g$ gives more weight to the tail than the identity function $g(x)=x$, whereas any convex distortion function $g$ gives less weight to the tail than the identity function $g(x)=x$.

\subsection {Distorted risk measures}

Let $(\Omega, \Bbb{F}, P)$ be a probability space on which all random variables involved are defined. Let $F_X$ be the cumulative distribution function of random variable $X$ and the decumulative distribution function
is denoted by $\bar{F}_X$, i.e. $\bar{F}_X(x)=1-F_X(x)=P(X>x)$.
 Let $g$ be a distortion function.
The distorted expectation of the random variable $X$, notation $\rho_{g}[X]$, is defined as
$$\rho_{g}[X]=\int_0^{+\infty}g(\bar{F}_X(x))dx+\int_{-\infty}^0 [g(\bar{F}_X(x))-1]dx,$$
provided at least one of the two integrals above is finite. If $X$ a non-negative  random variable, then $\rho_{g}$ reduces to
$$\rho_{g}[X]=\int_0^{+\infty}g(\bar{F}_X(x))dx.$$
From a mathematical point of view, a distortion  expectation
is the Choquet integral (see Denneberg (1994)) with respect to the nonadditive measure   $\mu=g\circ P$. That is  $\rho_{g}[X]=\int X d\mu$.
 In view of Dhaene et al. (2012, Theorems 4 and 6) we know  that, when the distortion function $g$ is right continuous
on $[0,1)$, then  $\rho_{g}[X]$ may be rewritten as
$$\rho_{g}[X]=\int_{[0,1]}VaR^+_{1-q}[X]dg(q),$$
where
$VaR^+p[X]=\sup\{x|F_X(x)\le p\}$, and when the distortion function $g$ is left continuous
on $(0,1]$, then  $\rho_{g}[X]$ may be rewritten as
$$\rho_{g}[X]=\int_{[0,1]}VaR_{1-q}[X]dg(q)=\int_{[0,1]}VaR_{q}[X]d{\bar g}(q),$$
where
$VaR_p[X]=\inf\{x|F_X(x)\ge p\}$ and ${\bar g}(q):=1-g(1-q)$ is the dual distortion of $g$. Obviously, ${\bar {\bar g}}=g$,
$g$ is left continuous if and only if ${\bar g}$ is right continuous; $g$ is concave  if and only if ${\bar g}$ is convex.
The distorted expectation $\rho_{g}[X]$ is called a
distortion risk measure with distortion function $g$. Distortion risk measures   are a particular class of risk measures
 which as premium principles were introduced by Deneberg
(1994) and further developed by Wang (1996, 2000) among others.
As it is well known,
the mathematical expectation, $E[X]$, is a distortion risk measure whose distortion function is
the identity function. If $g$ is concave, then
$$\rho_{g}[X]\ge\int_0^{+\infty}\bar{F}_X(x)dx+\int_{-\infty}^0 [\bar{F}_X(x)-1]dx=E[X],$$
and if $g$ is convex, then
$$\rho_{g}[X]\le\int_0^{+\infty}\bar{F}_X(x)dx+\int_{-\infty}^0 [\bar{F}_X(x)-1]dx=E[X].$$

 Distortion risk measures satisfy a set of properties including
positive homogeneity, translation invariance and monotonicity. Hardy and Wirch (2001) have shown that a risk measure based on a distortion
function is coherent if and only if the distortion function is concave.
A risk measure is said to be coherent if it satisfies the following set of four
properties (see, e.g., Arztner et al. 1997 and 1999):\\
(M) Monotonicity: $\rho(X)\le\rho(Y)$ provided that $P(X\le Y) = 1$.\\
(P) Positive homogeneity: For any  positive
constant $c>0$ and loss $X, \rho(cX) = c\rho(X)$.\\
(S) Subadditivity: For any losses $X, Y$, then
$\rho(X+Y)\le \rho(X)+\rho(Y)$.\\
(T) Translation invariance: If $c$ is a constant, then
$\rho(X+c)=\rho(X)+c$.\\
It is furthermore shown by Artzner et al. (1999) that all mappings satisfying the above properties allow a
representation:
$$\rho(X) =\sup_{p\in \mathcal{P}}E_p[X], $$
where $\mathcal{P}$ is a collection of  `generalised scenarios'.
A  risk measure $\rho$ is called a convex risk measure if it satisfies monotonicity, translation invariance
and the following convexity (C):
  $$ \rho(\lambda X+(1-\lambda)Y)\le \lambda \rho(X)+(1-\lambda)\rho(Y),\; 0\le \lambda\le 1.$$
Clearly, under the assumption of positive homogeneity, monotonicity and translation invariance, the convexity of a  risk measure is equivalent subadditivity.

The most well-known examples  of  distortion risk measures are the above-mentioned   VaR and TVaR,
  corresponding to the distortion functions, respectively, are
$g(x)={\bf 1}_{(x>1-p)}$ and $ g(x)=\min\left\{\frac{x}{1-p}, 1\right\}$.
 Notice that
${\rm TVaR}_p[X]$ can be alternatively expressed as the weighted average of VaR and losses exceeding VaR:
\begin{equation}
{\rm TVaR}_p[X]=VaR_p[X]+\frac{1-F_X(VaR_p[X])}{1-p} E\left[X-VaR_p[X]|X>VaR_p[X]\right].
\end{equation}
 For continuous distributions, TVaR coincide with the expected loss exceeding $p$-Value-at Risk, i.e., the mean of the worst $(1-p) 100\%$ losses in a specified time period
 which defined by
$$CTE_p[X]=E\left[X|X>VaR_p[X]\right].$$
If $X$ is a real valued random variable and $0 < p < 1$, then we say
that $q$ is an $p$-quantile if $P[X <q] \le p \le P[X \le q]$. By definition, ${\rm VaR}_p[X]$ is the lower $p$-quantile of the r.v. $X$ and ${\rm VaR}^+_p[X]$ is the upper $p$-quantile of the r.v. $X$. ${\rm VaR}_p[X]$  is a left-continuous nondecreasing function having ${\rm VaR}_0[X]$  as the essential infimum of $X$, possibly $-\infty$, ${\rm VaR}^+_p[X]$ is a right-continuous nondecreasing function having ${\rm VaR}_1[X]$  as the essential supermum of $X$, possibly $+\infty$.
  It is easy to see that ${\rm VaR}_p[X]\le {\rm VaR}^+_p[X]$, there are at most countably many values of $p\in [0,1]$ where ${\rm VaR}_p[X]$ and ${\rm VaR}^+_p[X]$ differ (see, Dhaene et al. (2012)). Moreover,
${\rm VaR}_p[X]={\rm VaR}^+_p[X]$ if, and only if $F_X(x)=p$ for at most one $x$, which equivalent to $F_X(\cdot)$
is strictly increasing.   The
risk measures VaR and VaR$^+$ satisfy axioms (M), (P), and (T), but not
(S) and (C) (except in some special cases, such as in the multivariate normal distributions or more generally multivariate elliptical  distributions), and hence is not coherent in the sense of Artzner et al. (1999).
 Despite  suffers from some serious limitations,  VaR is still the  standard of industry and regulatory  for the calculation of risk capital in banking and insurance. For example,  the Basel Committee on Banking Supervision introduced a 99\% Value at Risk requirement, based on a 10-day trading horizon.
The TVaR improves the VaR as a measure of risk by also taking into account the
magnitude of loss beyond the VaR. That is TVaR measures average losses in the most adverse cases
rather than just the minimum loss, as the VaR does. Therefore, risk
assessment based on the TVaR have to be considerably higher than those based on
VaR. The importance of TVaR is also seen from a
result of Kusuoka (2001), who proved that  $TVaR_p$ is the smallest law invariant coherent
risk measure that dominates $VaR_p$.
Unlike VaR, the distortion function associated to the  TVaR   is concave and, then, the  TVaR  is a coherent risk measure in the sense of Artzner et al. (1999). It means that TVaR  is a subadditive risk measure (see, for instance, Denuit et al., 2006). In the literature, the   TVaR  is sometimes called the expected shortfall.
Although  TVaR is  one of the best coherent risk measures, however, TVaR reflects only the mean size of losses exceeding the  VaR. It ignores the useful information in a large part of
the loss distribution, and consequently lacks incentive for mitigating losses below
the quartile VaR. Moreover, it does not properly adjust for extreme low-frequency
and high-severity losses, since it only accounts for the mean value (not higher
moments). A recent paper by Frittelli et al. (2014) has proposed a new risk measure, the lambda
value at risk $\Lambda$VaR) as a generalization of the VaR. The novelty of the $\Lambda$VaR lies in the fact that the
confidence level can change and adjust according to the risk factor profit
and loss.

Detailed studies of distortion risk measures and their relation with orderings of risk and the concept of comonotonicity
can be found in, for example, Wang (1996), Wang and Young (1998), H\"urlimann (1998),  Hua and Joe (2012) and the references therein.
The following lemma will be used
 in proofs of later results, which characterizes an
ordering of distortion risk measures in terms of their distortion functions.

\begin{lemma} (Belles-Sampera et al. (2014b)).   If  $g(x)\le g^*(x)$ for $x\in[0,1]$, then $\rho_{g}[X]\le \rho_{g^*}[X]$ for any random variable $X$.
\end{lemma}

\section {Generating new   distortion functions and measures}

Distortion functions can be considered as a starting point for constructing families of   distortion risk measures.
 Thus, constructions of  distortion functions play an
important role in producing various families of  risk measures.
Using the technique of  mixing, composition and copula  allow the construction of   new class of  distortion functions and measures.

\subsection {Composting methods }
The first approach to construct  distortion functions is the composition of distortion functions.

Let  $h_1, h_2,\cdots$ be distortion functions, define  $f_1(x)=h_1(x)$ and composite functions
$f_n(x)=f_{n-1}(h_n(x)), n=1,2,\cdots$. It is easy to check that $f_n(x), n=1,2,\cdots$ are all distortion functions.
If   $h_1, h_2,\cdots$ are  concave distortion functions, then each $f_n(x)$ is concave and satisfies that
 $$f_1\le f_2\le f_3\le \cdots$$ and
$$\lim_{n\to\infty}f_n(x)={\bf 1}_{(x>0)}, \; x\in[0,1].$$
The associated   risk measure  satisfies (by Lemma 2.1)
$$\rho_{f_1}[X]\le \rho_{f_2}[X]\le  \rho_{f_3}[X]\le \cdots$$ and
$$\lim_{n\to\infty}\rho_{f_n}[X]=VaR_1[X]={\rm esssup}(X).$$
  If   $h_1, h_2,\cdots$ are  convex distortion functions, then each $f_n(x)$ is convex and satisfies   that
 $$f_1\ge f_2\ge f_3\ge \cdots$$ and
$$\lim_{n\to\infty}f_n(x)={\bf 1}_{(x=1)}, \; x\in[0,1].$$
The associated   risk measure  satisfies (by Lemma 2.1)
$$\rho_{f_1}[X]\ge \rho_{f_2}[X]\ge  \rho_{f_3}[X]\ge \cdots$$ and
$$\lim_{n\to\infty}\rho_{f_n}[X]=VaR_0[X]={\rm essinf}(X).$$

 Consider   two distortion functions $g_1$ and $g_2$.
 If
 \begin{eqnarray*}
g_2(x)=\left\{  \begin{array}{ll} \frac{x}{1-p},  \ &{\rm if}\ 0\le x\le 1-p,\\
1, \ &{\rm if}\ 1-p< x\le 1, \end{array}
  \right. \nonumber
\end{eqnarray*}
then we get
 \begin{eqnarray*}
  g_p(x):=g_1(g_2(x))=\left\{  \begin{array}{ll} g_1\left(\frac{x}{1-p}\right),  \ &{\rm if}\ 0\le x\le 1-p,\\
1, \ &{\rm if}\ 1-p< x\le 1. \end{array}
  \right.
\end{eqnarray*}
The corresponding risk measure  $\rho_{g_p}[X]$ is the  tail distortion risk measure
which  was first introduced by Zhu and Li (2012), and was reformulated by Yang (2012). In particular, on the space of continuous loss random variables $X$,
$$\rho_{g_p}[X]=\int_0^{\infty}g_p\left(1-P(X\le x|X>VaR_p[X])\right)dx.$$

If $g_1(x)=x^r, 0<r<1$ and
\begin{eqnarray*}
g_2(x)=\left\{  \begin{array}{ll} \frac{x}{1-p},  \ &{\rm if}\ 0\le x\le 1-p,\\
1, \ &{\rm if}\ 1-p< x\le 1, \end{array}
  \right.
\end{eqnarray*}
then
\begin{eqnarray*}
  g_{12}(x):=g_1(g_2(x))=\left\{  \begin{array}{ll} \left(\frac{x}{1-p}\right)^r,  \ &{\rm if}\ 0\le x\le 1-p,\\
1, \ &{\rm if}\ 1-p< x\le 1, \end{array}
  \right.
\end{eqnarray*}
and
\begin{eqnarray*}
  g_{21}(x):=g_2(g_1(x))=\left\{  \begin{array}{ll} \frac{x^r}{1-p},  \ &{\rm if}\ 0\le x\le (1-p)^{\frac{1}{r}},\\
1, \ &{\rm if}\  (1-p)^{\frac{1}{r}}< x\le 1. \end{array}
  \right.
\end{eqnarray*}
Clearly, $g_1<g_{21}$ and $ g_2<g_{12}$, so that, by Lemma 2.1, $\rho_{g_1}[X]<\rho_{g_{21}}[X]$ and
$\rho_{g_2}[X]<\rho_{g_{12}}[X]$.

In practice, sometimes one  needs distort the initial distribution more than one times.

{\bf Example 3.1}\; Consider two risks $X$ and $Y$ with distributions, respectively, are:
\begin{eqnarray*}
F_X(x)=\left\{  \begin{array}{lll} 0,  \ &{\rm if}\ x<0,\\
0.6, \ &{\rm if}\ 0\le x<100,\\
0.975,\ &{\rm if}\ 100\le x<500,\\
1,\ &{\rm if}\ x\ge 500,\\
 \end{array}
  \right. \nonumber
\end{eqnarray*}
and
\begin{eqnarray*}
F_Y(x)=\left\{  \begin{array}{lll} 0,  \ &{\rm if}\ x<0,\\
0.6, \ &{\rm if}\ 0\le x<100,\\
0.99,\ &{\rm if}\ 100\le x<1100,\\
1,\ &{\rm if}\ x\ge 1100.\\
 \end{array}
  \right. \nonumber
\end{eqnarray*}
Then $EX=EY=50$,
VaR$_{0.95}$[X]=VaR$_{0.96}$[X]=100, VaR$_{0.95}$[Y]=VaR$_{0.96}$[Y]=100.\\
TVaR can be calculated by formula (2.1):\\
TVaR$_{0.95}$[X]=TVaR$_{0.95}$[Y]=300, TVaR$_{0.96}$[X]=TVaR$_{0.96}$[Y]=350.
So that when $\alpha=0.95$ and $\beta=0.96$, according to the measures of VaR and  TVaR,  both $X$ and $Y$  bear the same risk! However, the maximal loss for $Y$ (1100) is more than double than for loss $X$ (500), clearly, risk $Y$ is more risky than risk $X$.
Now we consider distortion expectation  $\rho_{g_p}$ with
\begin{eqnarray*}
g_1(x)=g_2(x)=\left\{  \begin{array}{ll} \frac{x}{1-p},  \ &{\rm if}\ 0\le x\le 1-p,\\
1, \ &{\rm if}\ 1-p< x\le 1. \end{array}
  \right.
\end{eqnarray*}
One can easily find that, with $p=0.95$, $\rho_{g_p}[X]=500$ and $\rho_{g_p}[Y]=1100$.

\subsection { Mixing methods}

One of the easiest ways to generate  distortion functions is to use the
method of mixing  along with finitely  distortion functions or  infinitely many distortion functions. Specifically,
if $g_w$ ($w\in <a,b>$) is a one-parameter family of distortion functions, $\psi$ is an increasing function on $<a,b>$
  such that $\int_{<a,b>}d\psi(w)=1$, then the function
$g=\int_{<a,b>}g_w d\psi(w)$ is a distortion function, the associated   risk
measure is  given by
\begin{equation}
\rho_{g}[X]=\int_{<a,b>}  \rho_{g_w} [X]d\psi(w).
\end{equation}
In particular, if $\psi$ is discrete distribution, then (3.1) can be written as the form of
 convex linear combination $g=\sum_i w_i g_i$ ($w_i\ge 0, \sum_i w_i=1)$ , the associated   risk
measure is  given by
\begin{equation}
\rho_{g}[X]=\sum_i w_i \rho_{g_i}[X].
\end{equation}

The following  lemma is well known (cf. Kriele and Wolf (2014, Theorem 2.1, P.33)).
\begin{lemma}\ If all $\rho_{g_w}$  ($w\in <a,b>$) are monotone, positively homogeneous, subadditive and translation invariant, then
$\rho_{g}[X]$ also has the corresponding properties. That is, if all $g_w$  ($w\in <a,b>$) are coherent, then $\rho_{g}[X]$ is also
coherent.
\end{lemma}

 Now we list three interesting special cases:

 $\bullet$ If $[a,b)=[0,\infty)$, $g_i(x)=1-(1-x)^i, i\ge 1$ and $w_i\ge 0, \sum_i w_i=1$, then  $\rho_{g}[X]$ in (3.2) is coherent since
 $g_i(x)$ is concave. As in Tsukahara (2009), if we take $w_i$ from Bin (1, $\theta)$ ($0<\theta<1$), then
 $g_{\theta}(u)=u+u\theta -u^2\theta.$  If we take
 $$w_i=\frac{\theta^i}{(e^{\theta}-1)i!},\; \theta>0,$$
 then
  $$g_{\theta}(u)=\frac{e^{\theta}(1-e^{-\theta u})}{e^{\theta}-1}.$$
 Also, if take  $w_i=(1-\theta)^{i-1}\theta$ ($0<\theta<1$), the geometric distribution, then
  $$g_{\theta}(u)=\frac{u}{u+\theta(1-u)},$$
  which is the proportional odds distortion; see    Example 2.1 in  Cherubini and Mulinacci (2014).

$\bullet$ If $[a,b]=[0,1]$, $\rho_{g_w}=VaR_w[X]$ and
$d\psi(w)=\phi(w)dw$, then $\rho_{g}[X]$ in (3.1) reduces to
\begin{equation}
\rho_{\phi}[X]=\int_0^1  VaR_w[X]\phi(w)dw,
\end{equation}
which is spectral risk measure (see  Acerbi 2002, 2004). Here $\phi$ is called a weighting function satisfies
the following properties:
 $\phi\ge 0$, $\int_0^1 \phi(w)dw=1$.  The following lemma gives a  sufficient condition for $\rho_{\phi}[X]$ to be a coherent  risk measure (cf. Kriele and Wolf (2014)).
\begin{lemma}\
Spectral risk measure  $\rho_{\phi}[X]$  is coherent if $\phi$ is (almost everywhere) monotone increasing.
\end{lemma}

 Clearly,  there exists a one-to-one
correspondence between distortion function $g$ and weighting function $\phi$,
namely, $g(1-t)=1-\int_0^t \phi(s)ds.$ Obviously, $g$ is concave if, and only if $\phi$
is (almost everywhere) monotone increasing.
Two well-known members of this class are the VaR and the TVaR.  The associated  weight functions
are $\phi(w)=\delta_p(w)$ and $\frac{1}{1-p}{\bf 1}(w>p)$, respectively. Here $\delta_p(w)$ is
a Dirac delta function that gives the outcome $\alpha=p$ an infinite weight and gives every other outcome a weight of zero.  From Lemma 3.2, TVaR is coherent since $\phi(w)=\frac{1}{1-p}{\bf 1}(w>p)$ is monotone increasing. By contrast,  $\phi(w)=\delta_p(w)$ is not  monotone increasing, hence VaR is not coherent.
 Both of these measures use only the tail of the distribution.

$\bullet$  If $[a,b]=[0,1]$, $\rho_{g_w}=TVaR_w[X]$ and
   $\psi=\mu$ is a probability measure on $[0,1]$,
  then  $\rho_{g}[X]$ in (3.1) reduces to
  \begin{equation}
\rho_{\mu}[X]=\int_0^1  TVaR_w[X]d\mu(w),
\end{equation}
which is  the weighted  TVaR (see  Cherny (2006)). TVaR$_p$ is a special weighted  TVaR with $\mu(w)={\bf 1}(w\ge p)$.
According to Lemma 3.1, since each  TVaR$_w[X]$  is coherent risk measure,  the weighted  TVaR is
coherent risk measure.
The weighted  TVaR can be rewritten as the form of spectral risk measure as following:
\begin{eqnarray*}
\rho_{\mu}[X]&=&\int_0^1  TVaR_w[X]d\mu(w)=\int_0^1\left(\frac{1}{1-w}\int_w^1 VaR_q[X] dq\right) d\mu(w)\\
&=&\int_0^1 \left(VaR_q[X]\int_0^q \frac{1}{1-w} d\mu(w)\right)dq \; ({\rm by\; the\;  Fubini \; theorem})\\
&=&\int_0^1 VaR_q[X] \phi(q)dq\\
&=&\int_0^1 VaR_{1-q}[X]dg(q),
\end{eqnarray*}
where, $g$ is a function with $g(0)=0$ and satisfies
$$ g'(1-q)=\phi(q)=\int_0^q \frac{1}{1-w} d\mu(w).$$  Because $\phi(q)$ is increasing function of $q$, it follows from Lemma 3.2 that the weighted  TVaR $\rho_{\mu}[X]$ is  coherent. Or, equivalently, $g'(q)$ is decreasing function of $q$, i.e. $g$ is a concave function, moreover,  $g$ is increasing and
\begin{eqnarray*}
g(1)&=&\int_0^1 g'(1-w)dw=\int_0^1 dq \int_0^q \frac{1}{1-w} d\mu(w)\\
&=&\int_0^1 \frac{1}{1-w} d\mu(w)\int_w^1 dq\\
&=&\int_0^1d\mu(w)=1.
\end{eqnarray*}
so that $g$ is a concave  distortion function, and hence
 the weighted  TVaR $\rho_{\mu}[X]$ is  coherent.

Conversely, the distortion measure with concave distortion function $g$ can be expressed  by the  weighted  TVaR. In fact, note that $\phi(q)=g'(1-q)$ is monotone increasing,  we define a measure $\nu([0,q])=\phi(q)$. As in the proof of Theorem 2.4 in Kriele and Wolf (2014) we have
\begin{eqnarray*}
\rho_{g}[X]&=&-\int_0^1  VaR_w[X] dg(1-w)\\
&=&\int_0^1  VaR_w[X] g'(1-w)dw\\
&=&\int_0^1  VaR_w[X] \phi(w)dw\\
&=&\nu([0, 1])EX+\int_0^1 TVaR_w[X](1-w)d\nu(w)\\
&=&\nu([0, 1])EX+\int_0^1 TVaR_w[X]d\mu(w),
\end{eqnarray*}
where
$$d\mu(w)=(1-w)d\nu(w).$$ It can be shown that $\mu$ is a probability measure. In fact,
\begin{eqnarray*}
\int_0^1 d\mu(w)&=&\int_0^1 \nu([0,w])dw\\
&=&\int_0^1 \phi(w)dw=\int_0^1 g'(w)dw=1.
\end{eqnarray*}

We now give some examples of interesting distortion functions and risk measures.

{\bf Example 3.2}\ If $w_1, w_2, w_3, w_4 \ge 0, \sum_{i=1}^4 w_i=1$, then
 $$g_{\alpha\beta}(x)=w_1 \nu_{\beta}(x)+w_2 \nu_{\alpha}(x)+w_3 \psi_{\beta}(x)+w_4 \psi_{\alpha}(x),$$
 is a distortion function,
where $\nu_{\beta}, \nu_{\alpha}, \psi_{\beta}, \psi_{\alpha}$ are the distortion functions of TVaR and VaR at confidence levels $\beta$ and $\alpha$, respectively. Then the corresponding
risk measure
$$\rho_{g_{\alpha\beta}}[X]=w_1 TVaR_{\beta}[X]+w_2 TVaR_{\alpha}[X]+w_3 VaR_{\beta}[X]+ w_4 VaR_{\alpha}[X],$$
 is called the GlueVaR risk measure, which were initially defined by
 Belles-Sampera et al. (2014a) (in the case $w_4=0$) and
the closed-form expressions of GlueVaR for Normal, Log-normal, Student's $t$ and Generalized
Pareto distributions  are provided. Two new proportional capital allocation principles based on GlueVaR risk
measures are studied in Belles-Sampera et al. (2014b).

Although  GlueVaR has superior mathematical properties than VaR and TVaR, however, the GlueVaR risk measure may also fails to recognize the differences between two  risks. For example,  consider two risks $X$ and $Y$ in Example 3.1,
we have computed that
VaR$_{0.95}$[X]=VaR$_{0.96}$[X]=100, VaR$_{0.95}$[Y]=VaR$_{0.96}$[Y]=100.
 TVaR$_{0.95}$[X]=TVaR$_{0.95}$[Y]=300, TVaR$_{0.96}$[X]=TVaR$_{0.96}$[Y]=350.
So that when $\alpha=0.95$ and $\beta=0.96$, we have $\rho_{g_{\alpha\beta}}[X]=\rho_{g_{\alpha\beta}}[Y]$.
 Thus  according to $\rho_{g_{\alpha\beta}}$,  both $X$ and $Y$  bear the same risk!
However, the maximal loss for $Y$ (1100) is more than double than for loss $X$ (500), clearly, risk $Y$ is more risky
than risk $X$.

{\bf Example 3.3}\ Let $\lambda\in [0,1]$, define a distortion function
$$g_{\lambda}(x)=\lambda g_0(x)+(1-\lambda) g(x),$$
where $g_0(x):={\bf 1}_{(x>0)}$ and $g$ is an arbitrary distortion function. Note that
$g_{\lambda}$ can be rewritten as
\begin{eqnarray*}
g_{\lambda}(x)=\left\{  \begin{array}{ll} 0,  \ &{\rm if}\ x=0,\\
\lambda+(1-\lambda)g(x), \ &{\rm if}\ 0< x\le 1. \end{array}
  \right.
\end{eqnarray*}
In particular, if $g(x)=x$, then  we get the esssup-expectation convex combination distortion
function with weight $\lambda$ on the essential supremum, which was introduced in Bann\"{o}r and Scherer (2014).
The corresponding risk measure
$$\rho_{g_{\lambda}}[X]=\lambda\ {\rm esssup} (X)+(1-\lambda)E(X),$$
which is a convex combination of the
essential supremum of $X$ and the ordinary expectation of $X$ w.r.t. $P$.

If  \begin{eqnarray*}
g(x)=\left\{\begin{array}{ll} [\alpha(1-\beta)+\frac{1-\alpha}{1-p}]x,  \ &{\rm if}\ 0\le x\le 1-p,\\
\alpha\beta+1-\alpha+\alpha(1-\beta)x, \ &{\rm if}\ 1-p< x\le 1, \end{array}
  \right. \nonumber
\end{eqnarray*}
where $0\le \alpha, \beta\le1$, $0<p<1$ are constants, then  we get
$$\rho_{g_{\lambda}}[X]=\lambda\ {\rm esssup} (X)+(1-\lambda)\rho_{g}[X],$$
where
$$\rho_{g}[X]=\alpha(1-\beta)E(X)+\alpha\beta VaR_p[X]+(1-\alpha)TVaR_p[X].$$

 As illustration, we consider the risks $X$ and $Y$ in Example 3.1, if $p=0.95$,
 then $\rho_{g}[X]=\rho_{g}[Y]=50\alpha\beta-250\alpha+300.$ It follows that
$$\rho_{g_{\lambda}}[X]=500\lambda+(1-\lambda)(50\alpha\beta-250\alpha+300)$$
 and
$$\rho_{g_{\lambda}}[Y]=1100\lambda+(1-\lambda)(50\alpha\beta-250\alpha+300).$$
Taking $\lambda=\frac12, \alpha=1, \beta=0$, then
$\rho_{g_{\lambda}}[X]=275$ and
 $\rho_{g_{\lambda}}[Y]=575$.
Taking $\lambda=\alpha=\beta=\frac12$, then
$\rho_{g_{\lambda}}[X]=437.5$ and $\rho_{g_{\lambda}}[Y]=737.5$. Thus the measure $\rho_{g_{\lambda}}$ can measure the differences between two  risks
$X$ and $Y$.

\subsection { A copula-based approach}

If $F$ is a distribution function on $[0,1]$, then $F$ can be used as a distortion function. The well-known examples are
 the PH transform and the dual power transform and, more generally,  the beta transform; see Wrich and Hardy (1999) for details.
 Similarly, we use this technique to a distribution function on $[0,1]^2$.
  We first  introduce the notion of copula in the two-dimensional case.

 {\bf Definition 3.1}. A two-dimensional copula $C(u,v)$ is a bivariate distribution on the square $[0,1]^2$ having uniform margins.
That is a function $C(u,v): [0,1]^2\rightarrow [0,1]$  is  right-continuous  in each variable such that
$C(0,v)=C(u,0)=0, C(u,1)=u, C(1,v)=v$ and  for $u_2>u_1, v_2>v_1$,
$$C(u_2,v_2)-C(u_2,v_1)-C(u_1,v_2)+C(u_1,v_1)\ge 0.$$
 For an introduction to copula theory and some of its applications, we refer to Joe (1997), Denuit et al. (2006) and
Nelsen (2006).

The well-known examples of copulas are $C^{+}(u, v) = \min(u, v), C^{\bot}(u, v) = uv$
and $C^-(u, v) = \max(u+v-1, 0)$ describing, respectively, comonotone dependence,
independence and countermonotone dependence between two random variables $X$ and $Y$. The copula version of the Fr\'{e}chet-Hoeffding
bounds inequality tells us
$$C^-(u, v)\le C(u, v)\le C^{+}(u, v).$$
Any copula  has the following decomposition (cf.  Yang et al (2006))
$$C(u, v) =\alpha C^+(u, v) + \beta C^{\bot}(u, v) +\gamma
C^{-}(u, v) + lG(u, v), $$
where $\alpha, \beta, \gamma, l\ge 0$, $\alpha+ \beta+ \gamma+ l=1.$  Here $G$ is a copula which called the indecomposable part.

For a given   two-dimensional copula $C(\cdot,\cdot)$, define one-parameter family $\{g_{p}\}_{p\in(0,1]}$ by
$g_{p} (u)=\frac{C(u,p)}{p}$ or $\frac{C(p,u)}{p}$. Clearly, for each $p$, $g_{p}$ is a right continuous distortion function. For example,

 $\bullet$ $g_{p} (u)=\frac{C^{\bot}(u,p)}{p}=u$  is continuous and  both  convex and concave, the associated
  risk measure  is $EX$;

$\bullet$  $g_{1-p} (u)=\frac{C^+(u,1-p)}{1-p}=\min\left\{\frac{x}{1-p},1\right\}$ is continuous and   concave, the corresponding risk measure is TVaR$_p$;

$\bullet$ $g_{p} (u)=\frac{C^-(u,p)}{p}=\max\left\{\frac{u+p-1}{p},0\right\}$ is continuous and   convex, the corresponding risk measure is
$\rho_{g_{p}}[X]=\frac{1}{p}\int_0^p VaR_q[X]dq$.

Conversely, if  $\{g_{p}\}_{p\in(0,1]}$   is a family of distortion functions, then, however,  $C(u, p)=p g_P(u)$ is not a copula in general;
 A sufficient condition can be found in  Cherubini and Mulinacci (2014).

A lot of copulas and methods to construct them can be found in the literature, for example, Joe (1997), Denuit et al. (2006) and Nelsen (2006).
We give below the most common bivariate copulas and the corresponding distortion functions.

$\bullet$ The Archimedean copulas:
$$C_{\Psi}(u,v)=\Psi^{[-1]}(\Psi(u)+\Psi(v))$$
for some generator $\Psi: (0,1]\rightarrow \Bbb{R}^+$ with $\Psi(1)=0$ such that $\Psi$ is   convex.
The pseudo-inverse of  $\Psi$ is the function
 $\Psi^{[-1]}$ with Dom$\Psi^{[-1]} = [0,\infty]$ and Ran$\Psi^{[-1]}$ = [0,1] given by
  \begin{eqnarray*}
 \Psi^{[-1]}(t)=\left\{\begin{array}{ll}  \Psi^{-1}(t),  \ &{\rm if}\ 0\le t\le \Psi(0),\\
 0, \ &{\rm if}\  \Psi(0)\le t\le\infty.
  \end{array}
  \right.
\end{eqnarray*}
If $\Psi$ is twice differentiable and $\Psi(0)=\infty$, then $C_{\Psi}$ is componentwise concave if, and only if $\frac{1}{\Psi'}$ is concave, where $\Psi'$ is the derivative of $\Psi$ (see Dolati and Nezhad (2014)). Aa a consequence, we have
\begin{theorem}
For each $v>0$, the  distortion function
$$g_{v}(u)=\frac{1}{v}\Psi^{-1}(\Psi(u)+\Psi(v))$$
  is concave if, and only if $\frac{1}{\Psi'}$ is concave.
\end{theorem}

We list some examples of  the Archimedean copulas and the corresponding distortion functions:

(a)\; The Clayton copula with parameter $\alpha>0$ is generated by  $\Psi(t)=\frac{1}{\alpha}(t^{-\alpha}-1)$
  and takes the form
$$C_{\alpha}(u,v)=(u^{-\alpha}+v^{-\alpha}-1)^{-1/\alpha},\; \alpha>0.$$
The limit of $C_{\alpha}(u,v)$  for $\alpha\downarrow  0 $ and $\alpha\uparrow\infty$ leads to independence and comonotonicity respectively (Nelsen, 2006).
The corresponding distortion functions:
$$g_{\alpha,v}(u)=\frac{1}{v}(u^{-\alpha}+v^{-\alpha}-1)^{-1/\alpha},\; \alpha>0, v\in (0,1].$$
In particular, if $\alpha=1$, we get the proportional odds distortion  which is found by  Cherubini and Mulinacci (2014):
 $$g_{1,v}(u)=\frac{u}{u+v(1-u)}.$$
 Since $(\frac{1}{\Psi'(t)})''=-\alpha(1+\alpha)t^{\alpha-1}<0$, $g_{\alpha,v}(u)$ is concave.

(b)\; In case $\Psi(t)=-\ln \left(\frac{e^{-\alpha t}-1}{e^{-\alpha }-1} \right)$ we get the Frank copulas:
$$C_{\alpha}(u,v)=-\frac{1}{\alpha}\ln \left(1+\frac{(e^{-\alpha u}-1)(e^{-\alpha v}-1)}{e^{-\alpha}-1}\right),\; \alpha \neq 0.$$
The corresponding distortion functions:
$$g_{\alpha,v}(u)=-\frac{1}{\alpha v}\ln \left(1+\frac{(e^{-\alpha u}-1)(e^{-\alpha v}-1)}{e^{-\alpha}-1}\right).$$
Since $(\frac{1}{\Psi'(t)})''=-\alpha e^{\alpha t}$, $g_{\alpha,v}(u)$  is  convex if $\alpha<0$ and  concave if $\alpha>0$.

(c)\; In case $\Psi(t)=t^{-1/\alpha}-1$ we get the Pareto survival  copulas:
$$\hat{C}_{\alpha}(u,v)=\left(u^{-1/\alpha}+v^{-1/\alpha}-1\right)^{-\alpha},\; \alpha>0.$$
The corresponding distortion functions:
$$g_{\alpha,v}(u)=\frac{1}{v}\left(u^{-1/\alpha}+v^{-1/\alpha}-1\right)^{-\alpha}.$$
Since  $(\frac{1}{\Psi'(t)})''=-(1+\frac{1}{\alpha})t^{\frac{1}{\alpha}-1}<0$, $g_{\alpha,v}(u)$ is concave.

(d)\; In case $\Psi(t)=\ln \frac{1+\alpha(t-1)}{t}, \alpha\in [-1,1)$ we get the Ali-Mikhail-Haq  copulas:
$$C_{\theta}(u,v)=\frac{uv}{1-\alpha(1-u)(1-v)}.$$
 The corresponding distortion functions:
$$g_{\theta,v}(u)=\frac{u}{1-\alpha(1-u)(1-v)}.$$
Since $(\frac{1}{\Psi'(t)})''=\frac{2\alpha}{\alpha-1}$, $g_{\theta,v}(u)$ is  convex if $\alpha\in [-1, 0]$ and  concave if $\alpha\in [0, 1)$.

 (e)\; In case $\Psi(t)=(-\ln t)^{\alpha}$ we get the Gumbel-Hougaard copulas:
$$C_{\alpha}(u,v)=\exp\left\{ -\left((-\ln u)^{\alpha}+(-\ln v)^{\alpha}\right)^{1/\alpha}\right\},\; \alpha\ge 1.$$
The corresponding distortion functions:
$$g_{\alpha,v}(u)=\frac{1}{v}\exp\left\{ -\left((-\ln u)^{\alpha}+(-\ln v)^{\alpha}\right)^{1/\alpha}\right\}.$$
The value  $\alpha=1$ gives independence and the limit for  $\alpha\uparrow\infty$
leads to comonotonicity.
Since \begin{eqnarray*}
 (\frac{1}{\Psi'(t)})''=\frac{\alpha-1}{\alpha t}(-\ln t)^{\alpha-2}\left(-1+\frac{2-\alpha}{\ln t}\right)
 \left\{ \begin{array}{lll} \le 0,  \ &{\rm if}\ 0<t\le 1, 1\le \alpha\le 2,\\
>0, \ &{\rm if}\ e^{2-\alpha}<t\le 1, \alpha>2,\\
<0,\ &{\rm if}\ 0<t<e^{2-\alpha}, \alpha>2,
 \end{array}
  \right.
\end{eqnarray*}
$g_{\alpha,v}(u)$ is  concave if $1\le\alpha\le 2$ and, if $\alpha>2$,  $g_{\alpha,v}(u)$ is convex on $(e^{2-\alpha},1]$ and concave on $(0, e^{2-\alpha})$.

Among other copulas, which do not belong to Archimedean family, it is worth to mention the following three copulas, given in  the bivariate case as:

 $\bullet$ The Farlie-Gumbel-Morgenstern copulas:
$$ C_{\alpha}(u,v)=uv+uv\alpha(1-u)(1-v),\; \alpha\in [-1,1],$$
The corresponding distortion functions:
$$g_{\alpha,v}(u)=u+u\alpha(1-u)(1-v),\; \alpha\in [-1,1], v\in[0,1],$$
which is  convex if $\alpha\in [-1, 0]$ and  concave if $\alpha\in [0, 1]$.

$\bullet$  The Marshall-Olkin copulas:
$$C_{\alpha,\beta}(u,v)=\min\{u^{1-\alpha}v, uv^{1-\beta}\},\; \alpha, \beta\in[0,1].$$
 Note that this copula is not symmetric for $\alpha\neq \beta$.
 The corresponding distortion functions:
$$g_{\alpha,\beta,v}(u)=\min\{u^{1-\alpha}, uv^{-\beta}\},\; \alpha, \beta\in[0,1], v\in(0,1],$$
which is  concave.  In particular, $g_{\alpha,0,v}(u)=g_{0,\beta,v}(u)=u$, $g_{1,1,v}(u)=\min\{1, \frac{u}{v}\}.$

$\bullet$  The normal copulas:
$$C_{\rho}(u,v)=\Phi_{\rho}(\Phi^{-1}(u),\Phi^{-1}(v)),$$
where  $\Phi_{\rho}$ is a bivariate normal distribution with standard normal marginal distributions and the correlation coefficient
$-1<\rho<1$, $\Phi^{-1}$ is the inverse function of the standard normal distribution.
The corresponding distortion functions:
$$g_{\rho,v}(u)=\frac{1}{v}\Phi_{\rho}(\Phi^{-1}(u),\Phi^{-1}(v)).$$

\section { Tail-asymptotics for VaR}

Subadditivity is an appealing property when aggregating risks in order to preserve the benefits of diversification.
Subadditivity of two risks is not only dependent on their dependence structure but also on the marginal distributions.
Value at risk is one of the most popular risk measures, but this risk measure is not always subadditive, nor convex,  exception of
 elliptically distributed risks. This family consists of many symmetric distributions such as the multivariate normal family,
 the multivariate Student-$t$ family,  the multivariate logistic family and  the multivariate exponential power family, and so on.
A recent development in the VaR literature concerns the  subadditivity in the tails (see Dan\'{i}elsson et al (2013)) which demonstrate that
VaR is subadditive in the tails of all fat tailed distributions, provided the
tails are not super fat.
However, in most practical models of interest the support of loss is bounded so that the maximum loss is
simply finite. We will also show that for this class losses VaR is subadditive  in the tail.
We can illustrate the ideas here with three simple  examples.  In Examples 4.1 and 4.3, $X$ and $Y$ are independent, while in
 Example 4.2, $X$ and $Y$ are dependent.

{\bf Example 4.1}\  Let $X$ and $Y$ be i.i.d. random variables which are Bernoulli (0.02) distributed, i.e.
$P(X=1)=1-P(X=0)=0.02$.
Then
$$P(X+Y=0)=P(X=0)P(Y=0)=0.98^2=0.9604,$$
$$P(X+Y=1)=P(X=1)P(Y=0)+P(X=0)P(Y=1)=0.0392,$$
$$P(X+Y=2)=P(X=1)P(Y=1)=0.0004.$$
Dhaene et al. (2006) verified that VaR is not subadditive since
VaR$_{0.975}[X]$=VaR$_{0.975}[Y]$=0 and VaR$_{0.975}[X+Y]$=1.
However, for $p\ge 0.98$, VaR$_{p}[X]$=VaR$_{p}[Y]$=1 and
\begin{eqnarray*}
 VaR_{p}[X+Y]=\left\{\begin{array}{ll} 1,  \ &{\rm if}\   0.98\le p<0.9996,\\
2, \ &{\rm if}\   p\ge 0.9996.
 \end{array}
  \right.
\end{eqnarray*}
Thus for $p\ge 0.98$,
$$ VaR_{p}[X+Y]\le VaR_{p}[X]+VaR_{p}[Y].$$

{\bf Example 4.2}\ Suppose we have losses $X$ and $Y$ , both dependent on the same underlying
Uniform(0,1) random variable $U$ as follows.
\begin{eqnarray*}
 X=\left\{\begin{array}{ll} 1000,  \ &{\rm if}\ U\le 0.04\\
0, \ &{\rm if}\   U>0.04
 \end{array}
  \right.
\end{eqnarray*}
\begin{eqnarray*}
 Y=\left\{\begin{array}{ll} 0,  \ &{\rm if}\ U\le 0.96\\
1000, \ &{\rm if}\   U>0.96
 \end{array}
  \right.
\end{eqnarray*}
Note that
$$P(X+Y=0)=P(X=0,Y=0)=P(U>0.04, U\le 0.96)=0.92,$$
$$P(X+Y=1000)=P(X=0,Y=1000)+P(X=1000,Y=0)=0.08.$$
Hardy (2006) found that VaR$_{0.95}[X]$=VaR$_{0.95}[Y]$=0,
VaR$_{0.95}[X+Y]$=1000. Thus
$$ VaR_{0.95}[X+Y]\ge VaR_{0.95}[X]+VaR_{0.95}[Y].$$
However, for any $\alpha>0.96$, VaR$_{\alpha}[X]$=VaR$_{\alpha}[Y]$=1000,
VaR$_{\alpha}[X+Y]=1000.$
Thus,
$$ VaR_{\alpha}[X+Y]\le VaR_{\alpha}[X]+VaR_{\alpha}[Y].$$

{\bf Example 4.3}\  Let $X$ and $Y$ be i.i.d. random variables which are  Uniform(0,1) distributed.
Then
\begin{eqnarray*}
 F_{X+Y}(z)=\left\{  \begin{array}{lll} 0,  \ &{\rm if}\ z<0,\\
\frac12 z^2, \ &{\rm if}\ 0\le z<1,\\
1-\frac12 (2-z)^2,\ &{\rm if}\ 1\le z<2,\\
1,\ &{\rm if}\ z\ge 2,\\
 \end{array}
  \right.
\end{eqnarray*}
and for $p\in (0,1]$, $VaR_p[X]=VaR_p[Y]=p$,
\begin{eqnarray*}
 VaR_p[X+Y]=\left\{\begin{array}{ll} \sqrt{2p},  \ & {\rm if}\ p\in (0,\frac12],\\
 2-\sqrt{2(1-p)},\ &{\rm if}  p\in [\frac12,1].
 \end{array}
  \right.
\end{eqnarray*}
 Thus for $p\in [\frac12,1]$,
$$ VaR_{\alpha}[X+Y]\le VaR_{\alpha}[X]+VaR_{\alpha}[Y].$$

Generally, we have the following result.

\begin{theorem} \  If the  risks $X_1, X_2,\cdots, X_k$ have finite upper endpoints,  then
$$\lim\sup_{p\to 1}\frac{VaR_p[\sum_{i=1}^k X_i]}{\sum_{i=1}^k VaR_p[X_i]}\le 1.$$
\end{theorem}
{\bf Proof}\ The proof is very simple. Denote by
${\rm esssup}(X_i)=\sup\{x: P(X_i\le x)<1\}$. Then ${\rm esssup}(X_i)<\infty$ and
$P(X_i\le {\rm esssup}(X_i))=1,    i=1,2,\cdots,k$, which lead to
$$P\left(\sum_{i=1}^k X_i\le \sum_{i=1}^k {\rm esssup}(X_i)\right)=1.$$
Hence
$${\rm esssup}\left(\sum_{i=1}^k X_i\right)\le \sum_{i=1}^k {\rm esssup}(X_i),$$
and the result follows.

\begin{remark}    Many distributions, such as Binomial, Uniform,   have finite upper endpoints;
Any truncated distribution: whether it is  right truncated or doubly
truncated all have finite upper endpoints.
\end{remark}
Next theorem consider the  random variables  $X_1, X_2, \cdots, X_k$ that are  not necessarily has finite upper endpoint, we first recall  the notion of (extended) regularly varying function:

\begin{definition} A function $f$ is called regularly varying at some point $x^-$ (or $x^+$, respectively) with index $\alpha\in \Bbb{R}$ if for all $t>0$,
$$\lim_{s\uparrow x}\frac{f(st)}{f(s)}=t^{\alpha}$$
(or $\lim_{s\downarrow x}\frac{f(st)}{f(s)}=t^{\alpha}$, respectively).
We write $f\in {\cal{R}}_{\alpha}^{x^{-}}$ ($f\in {\cal{R}}_{\alpha}^{x^{+}}$, respectively). For $\alpha=0$ we say $f$ is slowly varying; for
$\alpha=-\infty$  rapidly varying.
\end{definition}
\begin{definition}
Assume that $F$ is the distribution function of a nonnegative random. We
say $F$ belongs to the extended regular variation  class, if there are some $0 <\alpha\le \beta <\infty$ such that
$$s^{-\beta}\le \liminf_{x\to\infty}\frac{\overline{F}(sx)}{\overline{F}(x)}\le \limsup_{x\to\infty}\frac{\overline{F}(sx)}{\overline{F}(x)}\le s^{-\alpha}, \; {\rm for \;all}\;  s\ge 1,$$
or equivalently
$$s^{-\alpha}\le \liminf_{x\to\infty}\frac{\overline{F}(sx)}{\overline{F}(x)}\le \limsup_{x\to\infty}\frac{\overline{F}(sx)}{\overline{F}(x)}\le s^{-\beta}, \; {\rm for \;all}\;  0<s\le 1.$$
\end{definition}
We write $F\in ERV(-\alpha,-\beta)$.

A standard reference to the topic of  (extended) regular variation is Bingham et al. (1987) while main results are summarized by Embrechts et al. (1997).

\begin{theorem} \ We assume that  $X_1, X_2,\cdots, X_k$  have the same absolutely continuous  marginal distributions $F$ with infinite upper endpoint. \\
{\rm (1)}\; {\rm If}\;
  \begin{eqnarray}
  \lim_{z\to \infty}\frac{P(\sum_{i=1}^k X_i>z)}{P(X_1>\frac{z}{k})}<1,
  \end{eqnarray}
then
 \begin{eqnarray}
 \lim_{p\to 1}\frac{ VaR_p[ \sum_{i=1}^k X_i]}{\sum_{i=1}^k VaR_p[X_i]}< 1;
 \end{eqnarray}
 {\rm (2)}\; {\rm If}\;
 $$\lim_{z\to \infty}\frac{P(\sum_{i=1}^k X_i>z)}{P(X_1>\frac{z}{k})}=1,$$
then
$$\lim_{p\to 1}\frac{ VaR_p[\sum_{i=1}^k  X_i]}{\sum_{i=1}^k VaR_p[X_i]}=1;$$
 {\rm (3)}\; {\rm If}\;
 $$\lim_{z\to \infty}\frac{P(\sum_{i=1}^k X_i>z)}{P(X_1>\frac{z}{k})}>1,$$
then
$$\lim_{p\to 1}\frac{ VaR_p[\sum_{i=1}^k X_i]}{\sum_{i=1}^k VaR_p[X_i]}> 1.$$

\end{theorem}
{\bf Proof}\ We prove (1) only since  the other cases follow immediately in the same way.
 Because all the   marginal distributions are absolutely continuous, so we have
for any $p\in (0,1)$,
$$P(X_1>VaR_p[X_1])=P\left(\sum_{i=1}^k X_i>VaR_p\left[\sum_{i=1}^k X_i\right]\right)=1-p.$$
This, together with (4.1),  implies that
\begin{eqnarray}
  \lim_{p\to 1}\frac{P(X_1>VaR_p[X_1])}{P\left(X_1>\frac{1}{k}VaR_p[\sum_{i=1}^k X_i]\right)}<1.
  \end{eqnarray}
  The absolute continuity of $F$ implies that $\overline{F}$ is continuous and  strictly monotone  decreasing.
  Then from (4.3) we have
 $$ \lim_{p\to 1}\frac{ VaR_p[X_1]}{\frac{1}{k}VaR_p[\sum_{i=1}^k X_i]}> 1,$$
which is (4.2).
 This completes the proof.

{\bf Example 4.4}\; Suppose that each $X_i$ is regularly varying with index $-\alpha<0$. When the $X_i$ are mutually independent,
 it follows from ( Feller 1971, p. 279) that
$$\lim_{s\to\infty}\frac{P(\sum_{i=1}^k X_i>s)}{P(X_1>\frac{s}{k})}=\frac{k}{k^{\alpha}}.$$
Thus we get
 \begin{eqnarray*}
 \lim_{p\to 1}\frac{VaR_p[\sum_{i=1}^k X_i]}{ \sum_{i=1}^k VaR_p[X_i]}
\left\{  \begin{array}{lll}
<1, \ &{\rm if}\ \alpha>1,\\
=1, &{\rm if}\ \alpha=1,\\
>1, &{\rm if}\ \alpha<1.\\
\end{array}
  \right.
\end{eqnarray*}
Suppose that the $X_i$ are commonotonic, i.e. $P(X_1=\cdots= X_k)=1$, then
$$\lim_{s\to\infty}\frac{P(\sum_{i=1}^k X_i>s)}{P(X_1>\frac{s}{k})}=1.$$
So that in the  case $\alpha=1$ the result for the independent and the commonotonic case are the same.

The following result generalizes Theorem 10 in Jang and  Jho (2007) in which  all $Y_i$'s are assumed identically distributed.

\begin{theorem} Suppose $Y_1, \cdots, Y_k$ are nonnegative random variables (but
not necessarily independent or identically distributed.) If $Y_1$ has distribution
$F$ satisfying  $1-F(x)=x^{-\alpha}L(x), \alpha>0, x>0$,  where $L\in  $ ${\cal{R}}_{0}^{\infty}$ is slowly varying at infinity.
If $\frac{P(Y_i>x)}{\overline{F}(x)}\rightarrow c_i$ and $\frac{P(Y_i>x, Y_j>x)}{\overline{F}(x)}\rightarrow 0, i\neq j$,
as $x\to\infty$, $i, j=1,2,\cdots, k$, then
\begin{eqnarray*}
\lim_{p\to 1}\frac{VaR_p[\sum_{i=1}^k Y_i]}{\sum_{i=1}^k VaR_p[Y_i]}\left\{  \begin{array}{lll}
<1, \ &{\rm if}\ \alpha>1,\\
=1, &{\rm if}\ \alpha=1,\\
>1, &{\rm if}\ \alpha<1.\\
\end{array}
  \right.
\end{eqnarray*}
\end{theorem}
{\bf Proof}\ It follows from Lemma 2.1 in  Davis and Resnick (1996) that
$$\frac{P(\sum_{i=1}^k Y_i>x)}{1-F(x)}\rightarrow \sum_{i=1}^k c_i,\; as\;  x\to\infty.$$
This leads to
\begin{eqnarray}
  \lim_{p\to 1}\frac{P(\sum_{i=1}^k Y_i>VaR_p[\sum_{i=1}^k Y_i])}{P(Y_1>VaR_p[\sum_{i=1}^k Y_i])}=\sum_{i=1}^k c_i.
  \end{eqnarray}
Because
$$P(Y_1>VaR_p[Y_1])=P\left(\sum_{i=1}^k Y_i>VaR_p\left[\sum_{i=1}^k Y_i\right]\right)=1-p.$$
Thus from (4.4) that
\begin{eqnarray*}
  \lim_{p\to 1}\frac{P(Y_1>VaR_p[Y_1])}{P(Y_1>VaR_p[\sum_{i=1}^k Y_i])}=\sum_{i=1}^k c_i,
  \end{eqnarray*}
which is equivalent to
\begin{eqnarray*}
  \lim_{p\to 1}\frac{P(Y_1>VaR_p[Y_1])}{P\left(Y_1> (\sum_{i=1}^k c_i) ^{-\frac{1}{\alpha}} VaR_p[\sum_{i=1}^k Y_i]\right)}=1.
  \end{eqnarray*}
This implies that
\begin{eqnarray*}
  \lim_{p\to 1}\frac{VaR_p[Y_1]}{(\sum_{i=1}^k c_i) ^{-\frac{1}{\alpha}} VaR_p[\sum_{i=1}^k Y_i]}=1,
  \end{eqnarray*}
since $\overline{F}$ is continuous and  strictly monotone  decreasing.
   Note that $c_1=1$, $c_i^{1/\alpha} VaR_p[Y_1]\sim  VaR_p[Y_i]$ (as $p\to 1$) and
\begin{eqnarray*}
(\sum_{i=1}^k c_i)^{1/\alpha}\left\{  \begin{array}{lll}
<\sum_{i=1}^k c_i^{1/\alpha}, \ &{\rm if}\ \alpha>1,\\
= \sum_{i=1}^k c_i, &{\rm if}\ \alpha=1,\\
> \sum_{i=1}^k c_i^{1/\alpha}, &{\rm if}\ \alpha<1,
\end{array}
  \right.
\end{eqnarray*}
completing the proof.

\begin{remark}
The above result is obtained by Embrechts et al. (2009) for identically distributed and Archimedean copula  dependent $Y_i$'s.
However, our result can not obtained from their's due to the following fact:
The famous Farlie-Gumbel-Morgenstern family, does not belong to Archimedean family,
 which has the form
$$F(x,y)=F_1(x)F_2(y)(1+\alpha\overline{F_1}(x)\overline{F_2}(y))$$
where $F_1, F_2$ are two distributions and $\alpha\in[-1, 1]$ is a constant. When $F_1=F_2$, it satisfying
$\frac{\overline{F}(x,x)}{\overline{F_1}(x)}\rightarrow 0$ as $x\to\infty$.
\end{remark}

In the next theorem we consider the extended  regularly varying instead of  regularly varying.
\begin{theorem} Suppose $Y_1, \cdots, Y_k$ are nonnegative random variables  with the common identical distribution function $F$.
  If $F\in ERV(-\alpha,-\beta)$   and $\frac{P(Y_i>x, Y_j>x)}{\overline{F}(x)}\rightarrow 0, i\neq j$,
as $x\to\infty$, $i, j=1,2,\cdots, k$, then\\
(1)\; If $\beta<1$,
$$\limsup_{p\to 1}\frac{VaR_p[\sum_{i=1}^k Y_i]}{\sum_{i=1}^k VaR_p[Y_i]}>1;$$
(1)\; If $\alpha>1$,
$$\liminf_{p\to 1}\frac{VaR_p[\sum_{i=1}^k Y_i]}{\sum_{i=1}^k VaR_p[Y_i]}<1;$$
(1)\; If $\alpha=\beta=1$,
$$\lim_{p\to 1}\frac{VaR_p[\sum_{i=1}^k Y_i]}{\sum_{i=1}^k VaR_p[Y_i]}=1.$$
\end{theorem}

{\bf Proof}\ It follows from Lemma 2.2 in Zhang et al. (2009) that
$$\frac{P(\sum_{i=1}^k Y_i>x)}{1-F(x)}\rightarrow  k,\; as\;  x\to\infty.$$
This leads to
\begin{eqnarray*}
  \lim_{p\to 1}\frac{P(\sum_{i=1}^k Y_i>VaR_p[\sum_{i=1}^k Y_i])}{P(Y_1>VaR_p[\sum_{i=1}^k Y_i])}=k,
  \end{eqnarray*}
from which and using the same argument as  that in the proof of Theorem 4.3 leads to
\begin{eqnarray}
  \lim_{p\to 1}\frac{P(Y_1>VaR_p[Y_1])}{P(Y_1>VaR_p[\sum_{i=1}^k Y_i])}=k.
  \end{eqnarray}
If $\beta<1$, then
$$\limsup_{ p\to 1}\frac{\overline{F}(k^{-\frac{1}{\beta}} VaR_p[\sum_{i=1}^k Y_i] )}{(k^{-\frac{1}{\beta}})^{-\beta}\overline{F}(VaR_p[\sum_{i=1}^k Y_i])}\le 1.$$
This and (4.5) imply that
$$\limsup_{ p\to 1}\frac{\overline{F}(VaR_p[Y_1])}{\overline{F}(k^{-\frac{1}{\beta}} VaR_p[\sum_{i=1}^k Y_i] )}\ge 1.$$
It follows that
\begin{eqnarray}
  \limsup_{p\to 1}\frac{k^{\frac{1}{\beta}}VaR_p[Y_1]}{ VaR_p[\sum_{i=1}^k Y_i]}\le1.
  \end{eqnarray}
Thus
\begin{eqnarray*}
  \limsup_{p\to 1}\frac{kVaR_p[Y_1]}{ VaR_p[\sum_{i=1}^k Y_i]}< 1.
  \end{eqnarray*}
Similarly, if $\alpha>1$,
\begin{eqnarray}
  \liminf_{p\to 1}\frac{k^{\frac{1}{\alpha}}VaR_p[Y_1]}{ VaR_p[\sum_{i=1}^k Y_i]}\ge1
  \end{eqnarray}
and hence
\begin{eqnarray*}
  \liminf_{p\to 1}\frac{kVaR_p[Y_1]}{ VaR_p[\sum_{i=1}^k Y_i]}>1.
  \end{eqnarray*}
If $\alpha=\beta=1$, then by (4.6) and (4.7) one has
  \begin{eqnarray}
  \lim_{p\to 1}\frac{k VaR_p[Y_1]}{ VaR_p[\sum_{i=1}^k Y_i]}=1.
  \end{eqnarray}
  This ends the proof of Theorem 4.4.

To give applications of our results  we  employ extreme value theory techniques.
A distribution function $F$ (or the  rv $X$) is said to belong
to the Maximum Domain of Attraction (MDA)
 of the extreme value distribution $H$ if there exist constants $c_n>0, d_n\in \Bbb{R}$ such that $c_n^{-1}(\{\max(X_1,\cdots, X_n\}-d_n)\stackrel{d}{\rightarrow} H$. We write $X\in MDA(H)$ or $F\in MDA(H)$.
 According to the Fisher-Tippett theorem (see Theorem 3.2.3 in Embrechts et al. (1997))
$H$ belongs to one of the three standard extreme value distributions:
\begin{eqnarray*}
{\rm Frechet \; type}: \Phi_{\alpha}(x)&=&\left\{\begin{array}{ll}
0, {\rm if}\ x\le 0,\\
\exp\{-x^{-\alpha}\}, \ {\rm if}\ x>0,\\
 \end{array} \alpha>0.
\right.\\
{\rm Weibull \; type}: \Psi_{\alpha}(x)&=&\left\{\begin{array}{ll}
\exp\{-(-x)^{\alpha}\}, {\rm if}\ x\le 0,\\
1, \ {\rm if}\ x>0,\\
 \end{array} \alpha>0.
\right.\\
{\rm Gumbel\; type}: \Lambda(x)&=&\exp\{-e^{-x}\},\; x\in \Bbb{R}.
\end{eqnarray*}
Let $x_F$ denote the right-endpoint of the support of  $F$: $x_F=\inf\{x:F(x)=1\}$.
Then we have the following results (see  Embrechts et al. (1997), PP. 132-157).

$\bullet$ Fr\'{e}chet case: For some $\alpha>0$,  $F\in MDA(\Phi_{\alpha}) \Leftrightarrow \overline{F}\in  \cal{R}_{-\alpha}^{\infty}.$\\
Examples are Pareto, Cauchy, Burr, Loggamma and Stable with index $\beta<2$.

$\bullet$ Weibull case: For some $\alpha>0$,  $F\in MDA(\Psi_{\alpha}) \Leftrightarrow x_F<\infty, \overline{F}(x_F-1/x)\in  \cal{R}_{-\alpha}^{\infty}.$\\
Examples are  Uniform and Beta distribution.

$\bullet$ Gumbel case:    $F\in MDA(\Lambda_{\alpha}) \Leftrightarrow x_F\le\infty$ and there exists a positive measurable function $a$ such that for $t\in\Bbb{R}$
\begin{equation}
\lim_{x\uparrow x_F}\frac{\overline{F}(x+t a(x))}{\overline{F}(x)}=e^{-t}.
\end{equation}
Examples are Exponential-like, Weibull-like, Gamma, Normal, Lognormal, Benktander-type-I and Benktander-type-II.
\begin{remark}

(1).\; For  $\alpha>0$, if   $X_1, X_2,\cdots, X_k\in MDA(\Psi_{\alpha})$,  in view of  Weibull case above  they are all have finite supports. It follows from Theorem 4.1, VaR$_p$ is subadditive for $p$ is sufficiently close to 1.

(2).\;  For  $\alpha>0$, if   $X_1, X_2,\cdots, X_k\in MDA(\Phi_{\alpha})$ and are identically distributed, $(-X_1, -X_2,\cdots, -X_k)$ has an Archimedean copula with generator $\psi$, which is regularly varying at $0$ with index $-\beta<0$. We apply (2.2) in Alink et al. (2004) and Definition 4.1 to obtain
\begin{eqnarray*}
\lim_{z\to \infty}\frac{P(\sum_{i=1}^k X_i>z)}{P(X_1>\frac{z}{k})}&=&\lim_{z\to \infty}\frac{P( \sum_{i=1}^k X_i>z)}{P(X_1>z)}
\frac {P(X_1>z)}{P(X_1>\frac{z}{k})}\\
&=&q_k(\beta,\alpha)\lim_{z\to \infty}\frac {P(X_1>z)}{P(X_1>\frac{z}{k})}\\
&=&\lim_{z\to \infty}\frac {P(X_1>z(q_k(\beta,\alpha))^{-1/\alpha})}{P(X_1>\frac{z}{k})}\\
&=& k^{-\alpha}q_k(\beta,\alpha)\left\{  \begin{array}{lll}
<1, \ &{\rm if}\ \alpha>1,\\
=1, &{\rm if}\ \alpha=1,\\
>1, &{\rm if}\ \alpha<1,
\end{array}
  \right.
\end{eqnarray*}
where in the last step we have used Lemma 3.1(d) in Embrechts et al. (2009) which states that
$$\min\{k^{\alpha},k\}\le q_k(\beta,\alpha)\le \max\{k^{\alpha},k\}.$$
\end{remark}
This, together with Theorem 4.2 we recover the result Theorem 2.5 in Embrechts et al. (2009).

(3).\; If   $X_1, X_2,\cdots, X_k\in MDA(\Lambda_{\alpha})$ have common distribution  $F$, $(-X_1, -X_2,\cdots, -X_k)$ has an Archimedean copula with generator $\psi$, which is regularly varying at $0$ with index $-\beta<0$. We apply (2.6) in  Alink et al. (2004)    to obtain
\begin{eqnarray*}\lim_{z\to \infty}\frac{P(\sum_{i=1}^k X_i>z)}{P(X_1>\frac{z}{k})}=e^{-\frac{1}{k}}q_k^G(\beta),
\end{eqnarray*}
where
$$q_k^G(\beta)=\int_{\sum_{i=1}^k x_i\le 1}\frac{d^k}{dx_1\cdots dx_k}\left(\sum_{i=1}^k e^{-\beta x_i}\right)^{-1/\beta}dx_1\cdots dx_k.$$
The constant $q_k^G(\beta)\le e^{\frac{1}{k}}$  is strictly increasing in $\beta$ with
$$\lim_{\beta\to 0}q_k^G(\beta)=0,\;\; \lim_{\beta\to \infty}q_k^G(\beta)=e^{\frac{1}{k}}.$$
For more details, see  Alink et al. (2004) for  the case $k=2$ and Chen et al. (2012) for general case.
Thus by Theorem 4.2,
\begin{eqnarray*}
 \lim_{p\to 1}\frac{ VaR_p[ \sum_{i=1}^k X_i]}{\sum_{i=1}^k VaR_p[X_i]}\le 1.
 \end{eqnarray*}
In particular, when $\alpha\to\infty$,
\begin{eqnarray*}
 \lim_{p\to 1}\frac{ VaR_p[ \sum_{i=1}^k X_i]}{\sum_{i=1}^k VaR_p[X_i]}= 1
 \end{eqnarray*}
\begin{remark}
Note that ${\rm VaR}_p[X]$  is a left-continuous nondecreasing function having ${\rm VaR}_0[X]$  as the essential infimum of $X$ and ${\rm VaR}_1[X]$  as the essential supermum of $X$. Thus under the assumptions of Theorem 4.1 or Theorem 4.2, if $p$  close to 1, we have
 $$VaR_p[X_1+X_2]\le VaR_p[X_1]+VaR_p[X_2],$$
 which, together with the positive homogeneity of ${\rm VaR}_p[X]$, implies that, if $p$  close to 1,
 the convexity is holds:
 $${\rm VaR}_p[\lambda X+(1-\lambda)Y]\le \lambda  {\rm VaR}_p[X]+(1-\lambda) {\rm VaR}_p[Y],\; 0\le \lambda\le 1.$$
\end{remark}
 From above analysis we see that, although, in general the  VaR risk measure lack of subadditivity and convexity. However,  one should not too worries about violations of subadditivity  for risk management applications relying on VaR, since in most practical circumstances it is subadditive, at least is subadditive in the tail, and the failure to be subadditive in a few situations is not sufficiently important to reject the VaR risk measure.

 \section { Tail-subadditivity for  distortion risk measures}

The tail-subadditivity property for GlueVaR risk measures were initially defined by
Belles-Sampera et al. (2014a) and the milder condition
of subadditivity in the tail region is investigated. Furthermore, they verified that
a GlueVaR risk measure is tail-subadditive if its associated distortion
function $k^{h1,h2}_{\beta,\alpha}(u)$ is concave in $[0, 1-\alpha)$, where  parameters $\alpha$ is confidence level and  $\beta$ is an extra confidence level such that
$0\le\alpha\le\beta\le1$ and,
\begin{eqnarray*}
 k^{h1,h2}_{\beta,\alpha}(u)=\left\{ \begin{array}{lll}
\frac{h_1}{1-\beta}u, \ &{\rm if}\ 0\le u<1-\beta,\\
h_1+\frac{h_2-h_1}{\beta-\alpha}(u-1+\beta), &{\rm if}\ 1-\beta\le u<1-\alpha,\\
1, &{\rm if}\ 1-\alpha\le u\le 1,
\end{array}
  \right.
\end{eqnarray*}
where $h_1$ and $h_2$ are two distorted survival probabilities  at levels $1-\beta$
and $1-\alpha$, respectively. Here $ 0\le h_1\le h_2\le 1 $.
 We note, however, from their proof to Theorem 6.1 that the result will
hold for any distortion function that is concave in $[0, 1-\alpha)$, not restricted to $k^{h1,h2}_
{\beta,\alpha}(u)$. In this section we state the result and give an alternative  proof. As in  Belles-Sampera et al. (2014a), for a given confidence level $\alpha$, the tail region of a
random variable $Z$ is defined as
$\Bbb{Q}_{\alpha,Z}=\{w|Z(w)>s_{\alpha}\}\subseteq \Omega$, where
$s_{\alpha}=\inf\{z|\overline{F}_Z(z)\le 1-\alpha\}$ is the $\alpha$-quantile. For simplicity, we use the notation
$S_Z(z):= \overline{F}_Z(z)$.
\begin{theorem} \ For a confidence level $\alpha\in [0,1]$ and  two risks $X, Y$   defined on the same probability space. If $\Bbb{Q}_{\alpha,X}\cap \Bbb{Q}_{\alpha,Y}\cap \Bbb{Q}_{\alpha,X+Y}\neq\emptyset$
and $g$ is a concave distortion function in $[0,1-\alpha)$,
then the distortion risk measure $\rho_{g}$ is tail-subadditive. That is
\begin{eqnarray*}
\int^0_{0\wedge m_{\alpha}}[g(S_{X+Y}(z))-1]dz&+&\int^{\infty}_{0\vee m_{\alpha}} g(S_{X+Y}(z))dz\\
&\le & \int^0_{0\wedge m_{\alpha}} [g(S_{X}(z))-1]dz+\int^{\infty}_{0 \vee m_{\alpha}} g(S_{X}(z))dz\\
&&+\int^0_{0\wedge m_{\alpha}} [g(S_{Y}(z))-1]dz+\int^{\infty}_{0 \vee m_{\alpha}} g(S_{Y}(z))dz,
\end{eqnarray*}
where
$m_{\alpha}=\sup\{ s_{\alpha}(X),  s_{\alpha}(Y),  s_{\alpha}(X+Y)\}.$
\end{theorem}

{\bf Proof}\ Without loss of the generality, we assume that the risks $X$ and $Y$ are nonnegative, so that $m_{\alpha}=S^{-1}_{X+Y}(1-\alpha)\ge 0$.
It follows that
\begin{eqnarray*}
\int^{\infty}_{m_{\alpha}} g(S_{X+Y}(z))dz&=&\int_{S^{-1}_{X+Y}(1-\alpha)}^{\infty}  g(S_{X+Y}(z))dz\\
&=&\int_{S^{-1}_{X+Y}(1-\alpha)}^{\infty}dx\int_{[0,S_{X+Y}(x)]}dg(q)\\
&=&\int_{[0, S_{X+Y}(S^{-1}_{X+Y}(1-\alpha)))} dg(q)\int_{S^{-1}_{X+Y}(1-\alpha)}^{F_{X+Y}^{-1}(1-q)}dx\\
&=&\int_{[0, S_{X+Y}(S^{-1}_{X+Y}(1-\alpha)))}F_{X+Y}^{-1}(1-q) dg(q)-S^{-1}_{X+Y}(1-\alpha)g(1-\alpha),
\end{eqnarray*}
where in the third step we have used the Fubini's theorem to change the order of integration.
As above, we have
\begin{eqnarray*}
\int_{[0, S_{X+Y}(S^{-1}_{X+Y}(1-\alpha)))}F_{X+Y}^{-1}(1-q) dg(q)&=&\int_{1-S_{X+Y}(S^{-1}_{X+Y}(1-\alpha))}^1F_{X+Y}^{-1}(q) g'(1-q)dq\\
&=&\int_0^1 TVaR_{X+Y}(w)d\mu_{X+Y}(w),
\end{eqnarray*}
where
$$d\mu_{X+Y}(w)=(1-w)d\nu_{X+Y}(w), \;\;\nu_{X+Y}([0,q])={\bf 1}_{(1- S_{X+Y}(S^{-1}_{X+Y}(1-\alpha)),1]}(q)g'(1-q).$$
Finally, we get
 \begin{eqnarray}
 \int^{\infty}_{m_{\alpha}} g(S_{X+Y}(z))dz=\int_0^1 TVaR_{X+Y}(w)d\mu_{X+Y}(w)-S^{-1}_{X+Y}(1-\alpha)g(1-\alpha). \end{eqnarray}
Similarly,
 \begin{eqnarray}
 \int^{\infty}_{m_{\alpha}} g(S_{X}(z))dz=\int_0^1 TVaR_{X}(w)d\mu_{X}(w)-S^{-1}_{X+Y}(1-\alpha)g(1-\alpha), \end{eqnarray}
and
 \begin{eqnarray}
 \int^{\infty}_{m_{\alpha}} g(S_{Y}(z))dz=\int_0^1 TVaR_{Y}(w)d\mu_{Y}(w)-S^{-1}_{X+Y}(1-\alpha)g(1-\alpha), \end{eqnarray}
where
$$d\mu_{X}(w)=(1-w)d\nu_{X}(w),\; \nu_{X}([0,q])={\bf 1}_{(1- S_{X}(S^{-1}_{X+Y}(1-\alpha)),1]}(q)g'(1-q),$$
and
$$d\mu_{Y}(w)=(1-w)d\nu_{Y}(w),\; \nu_{Y}([0,q])={\bf 1}_{(1- S_{Y}(S^{-1}_{X+Y}(1-\alpha)),1]}(q)g'(1-q).$$
By the subadditivity of TVaR and note that $\nu_{X+Y}([0,q])\le \nu_{X}([0,q]), \nu_{Y}([0,q])$,
we obtain
\begin{eqnarray*}
\int_0^1 TVaR_{X+Y}(w)d\mu_{X+Y}(w)&\le &\int_0^1 TVaR_w(X)d\mu_{X+Y}(w)\\
&&+\int_0^1 TVaR_w(X)d\mu_{X+Y}(w)\\
&\le&\int_0^1 TVaR_{X}(w)d\mu_{X}(w)\\
&&+\int_0^1 TVaR_{Y}(w)d\mu_{Y}(w),
\end{eqnarray*}
this, together with (5.1)-(5.3), implies that
$$\int^{\infty}_{m_{\alpha}} g(S_{X+Y}(z))dz\le \int^{\infty}_{m_{\alpha}} g(S_{X}(z))dz+\int^{\infty}_{m_{\alpha}} g(S_{Y}(z))dz,$$
as desired.
\begin{remark}
Consider the distortion functions associated with the Gumbel-Hougaard copulas (cf. Section 3.3):
$$g_{\alpha,v}(u)=\frac{1}{v}\exp\left\{ -\left((-\ln u)^{\alpha}+(-\ln v)^{\alpha}\right)^{1/\alpha}\right\}.$$
 If $\alpha>2$, then $g_{\alpha,v}(u)$ is concave on $(0, e^{2-\alpha})$ and convex on $(e^{2-\alpha},1]$.
 Thus the distortion risk measure $\rho_{g_{\alpha,v}(u)}$ is tail-subadditive.
\end{remark}

\noindent {\bf Compliance with ethical standards}

\noindent {\bf Conflict of interest} \; The authors declare that they have no conflict of interest.

\noindent {\bf Human/Animals participants}\; This research does not involve human participants or animals.

\noindent{\bf Acknowledgements} \ 
The research   was supported by the National
Natural Science Foundation of China (11171179, 11571198) and the Research
Fund for the Doctoral Program of Higher Education of China (20133705110002).

\end{document}